\documentclass{ws-p9-75x6-50}

\begin{document}

\title{Approximate Solutions to the
Binary Black Hole Initial Value Problem}

\author{Pedro Marronetti}

\address{Center for Relativity, The University of Texas at Austin,
Austin Tx 78712, USA\\E-mail: pmarrone@physics.utexas.edu}

\maketitle

\abstracts{
We present approximate analytical solutions to the 
Hamiltonian and momentum constraint equations, corresponding
to systems composed of two black holes with arbitrary
linear and angular momentum. The analytical nature of these
initial data solutions makes them easier to implement in 
numerical evolutions than the traditional numerical approach
of solving the elliptic equations derived from the Einstein
constraints.}

Our work is based on descriptions of black holes in ingoing 
Eddington-Finkelstein (iEF) coordinates. This choice
is motivated by the fact that, in these coordinates, surfaces of
constant time ``penetrate'' the event horizon.
Foliations that penetrate the horizon facilitate the excision of
the singularity from the computational domain.  The essence of black
hole excision is the removal of the singularity while preserving the
integrity of the spacetime accessible to observers outside the black
hole.

Matzner, Huq, and Shoemaker \cite{Matzner:1998} presented a simple way
to construct approximate initial data for multiple black-hole systems,
which consists essentially in superposing the spatial metric corresponding to
single black holes with arbitrary spins, positions, and velocities.
In order to further reduce the residual errors, Marronetti {\it et al.} 
\cite{Marronetti:2000a} proposed variation of the superposition method 
that preserves the simplicity of being analytical. Essentially, the method 
consists of multiplying ``attenuation" functions into the recipe of the 
previous section. The purpose of the attenuation functions is to cancel 
the effects of a given hole on the neighborhood of the other hole,
to eliminate the singularities present in the constraint equations.

\begin{center}
\begin{figure}[t]
\epsfxsize=20pc 
\epsfbox{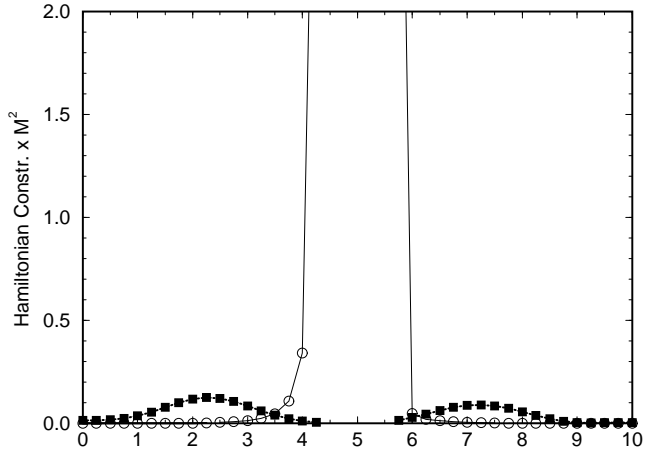} 
\caption{
Hamiltonian constraint residuals near the rightmost hole of the head-on 
collision.
\label{figure1}}
\end{figure}
\end{center}

We show here that even for interestingly close separation scenarios, 
the attenuation method errors are small, and in fact both the $l_1$ and 
$l_\infty$ norms are smaller than those of the truncation error 
(the discretization error in the calculation) in simulations at currently 
accessible computational resolutions. The results shown in this paper 
correspond to a head-on collision (which is simpler to display), but very 
similar results are found at comparable separations for non head-on data. 
The parameters of the system are 
$M_1 = M_2 = M = 1$, $|{\bf a_1}| = |{\bf a_2}| = a = 0.5M$ along the $z$ axis, 
and the holes are boosted against each other in the $x$ direction with velocity 
$v=0.5$. In order to evaluate the accuracy of these approximations, we 
calculate the residuals of the Hamiltonian and momentum constraint, since
they represent the departure from an exact initial data set 
(see \cite{Marronetti:2000a} for their definitions).
In this example, we excise the points inside a sphere of radius 
$a+h$ centered at the hole, where the absolute value of the specific angular 
momentum $a$ is also the radius of the Kerr ring-like singularity and $h$ is 
the grid spacing (for our example, $h=M/4$).
Figure \ref{figure1} shows the Hamiltonian constraint residuals near
the rightmost hole of the head-on collision. The circles show the second
order truncation error and the squares the residuals from the attenuation
method.
First we note that in some areas the violation of the constraints for 
the attenuated data is greater than that corresponding to truncation error. 
However, even in these areas the violation is smooth and small in absolute 
value, as opposed to the unbounded behavior shown by the truncation error 
near the singularity: the divergent behavior present at the singularity has 
disappeared with the use of attenuated data. This is due to the fact that, 
having cancelled the influence of the presence of the second hole, the 
metric and extrinsic curvature become the fields corresponding to 
an isolated black hole at the location of the singularity.
This simplifies the exact problem, since this method provides exact inner 
boundary conditions for the elliptic solver. The applicability of these
concepts to the numerical solution of the initial value problem 
has been confirmed by Marronetti and Matzner \cite{Marronetti:2000b},
who presented the first results for a black-hole binary
using the attenuation method to generate background data.

These methods can be further refined to provide more astrophysically
realistic initial data for the case of black holes in circular orbits, by
deriving the extrinsic curvature $K_{ij}$ from the presence of a Killing
vector $\xi =  \partial_t + \omega \partial_{\phi}$, instead of using
the superposition of two boosted holes. Work under progress will incorporate 
these variations into the full numerical problem.

\end{document}